\documentstyle[epsfig,epsf,rotate,a4,12pt]{article}
\epsfverbosetrue

\def\nuebar{\bar{\nu_e}}

\def\munu{\mu_{\nu}}
\def\munue{\mu_{\bar{\nu_e}}}
\def\mub{\rm{\mu_{B}}}

\def\dm2{\rm{\Delta m^2}}

\def\s2tw{\rm{ sin ^2 \theta _W }}
\def\am241{\rm{ ^{241} Am }}
\def\u238{\rm{ ^{238} U }}
\def\th232{\rm{ ^{232} Th }}
\def\k40{\rm{ ^{40} K }}
\def\enu{\rm{E_{\nu}}}

\def\munuebar{\rm{\mu_{\nuebar}}}

\begin{document}

\begin{flushright}
   {\bf AS-TEXONO/02-05}\\
        \today 
\end{flushright}

\begin{center}
\Large
\bf{
Status on the Searches of Neutrino Magnetic
Moment at the Kuo-Sheng Power Reactor
}
\end{center}

\begin{center}
\large
Henry Tsz-King Wong\footnote{Email: htwong@phys.sinica.edu.tw} \\
( on behalf of the TEXONO\footnote{{\bf T}aiwan 
{\bf EX}periment {\bf O}n {\bf N}eutrin{\bf O} :
Home Page at
http://hepmail.phys.sinica.edu.tw/$\sim$texono/
}
Collaboration )\\
\vspace*{0.1cm}
Institute of Physics, Academia Sinica, Taipei 11529, Taiwan\\
\end{center}


\vspace*{0.5cm}

\begin{center}
\large
{\bf Abstract}
\normalsize
\end{center}

The TEXONO collaboration has been built up among scientists
from Taiwan and China to pursue an experimental program
in neutrino and astro-particle physics. The flagship
efforts have been the study of low energy neutrino physics
at the Kuo-Sheng Power Reactor Plant in Taiwan. The
Reactor Laboratory is equipped with flexibly-designed shieldings,
cosmic veto systems, electronics and data acquisition
systems which can function with different detector schemes.
Data are taken during the Reactor Period June-01 till April-02 
with a high purity germanium detector and
46 kg of CsI(Tl) crystal scintillator array operating in parallel.
A threshold of 5 keV has been achieved for the germanium detector,
and the background level comparable to those of Dark Matter
experiments underground is achieved. 
Based on 62/46 days of analyzed Reactor ON/OFF data,
a preliminary result of 
$\rm{ ( \munue / 10^{-10} \mub )^2  =  - 1.1 \pm 2.5  }$
can be derived  for neutrino magnetic moment $\munue$.
Sensitivity region on neutrino radiative decay lifetime
is inferred. The complete data set would include
180/60 days of ON/OFF data.

\vfill

\begin{center}
\it{
( Contributed Paper to the International Conference
on High Energy Physics, 2002 )
}
\end{center}

\clearpage

\section{Introduction and History}

The 
TEXONO
Collaboration has been built up since 1997 to
initiate and pursue an experimental
program in Neutrino and Astroparticle Physics~\cite{texono}.
The Collaboration comprises
more than 40 research scientists from
major institutes/universities
in Taiwan (Academia Sinica$^{\dagger}$, Chung-Kuo
Institute of Technology, Institute of Nuclear
Energy Research, National Taiwan University, National Tsing Hua
University, and Kuo-Sheng Nuclear Power Station),
China (Institute of High Energy Physics$^{\dagger}$,
Institute of Atomic Energy$^{\dagger}$,
Institute of Radiation Protection,
Nanjing University, Tsing Hua University)
and the United States (University of Maryland),
with AS, IHEP and IAE (with $^{\dagger}$)
being the leading groups.
It is the first research collaboration 
of this size and magnitude
among major research institutes from Taiwan and China.

The research program~\cite{texono} emphasizes on the
the unexplored and unexploited theme of adopting
detectors with
high-Z nuclei, such as solid state device and
scintillating crystals,
for low-energy low-background experiments
in Neutrino and Astroparticle Physics\cite{prospects}.
The ``Flagship'' program~\cite{ksexpt} is 
a reactor neutrino experiment
to study low energy neutrino
properties and interactions
at the Kuo-Sheng (KS) Neutrino Laboratory.
It is the first particle physics experiment
performed in Taiwan.
In parallel to the reactor experiment,
various R\&D efforts coherent with the
theme are initiated and pursued. 

This article focuses on the magnetic moment data
taking and analysis
with a germanium detector during the Reactor Period
June 2001 till April 2002.

\section{Electromagnetic Properties of the Neutrinos}

The strong and positive evidence of neutrino oscillations
implies the existence of neutrino masses and mixings~\cite{pdg,ichep02},
the physical origin, structures and
experimental consequences of which
are still not thoroughly known and understood.
Experimental studies on the neutrino properties
and interactions which may reveal some of these
fundamental questions and/or constrain certain
classes of models are therefore of interests.
The coupling of neutrino with the photons are consequences of
non-zero neutrino masses.
Two of the manifestations
of the finite electromagnetic form factors
for neutrino interactions~\cite{vogelengel,nieves}
are neutrino magnetic moments and radiative decays.

The searches of neutrino magnetic moments
are performed in experiments on
neutrino-electron scatterings~\cite{kaiser}:
\begin{equation}
\rm{
 \nu_{\it l_1}   +    e^-   \rightarrow   \nu_{\it l_2}   +   e^-  .
  }
\end{equation}
The experimental observable is the kinetic energy of the
recoil electrons(T).
The differential cross section
for the magnetic scattering (MS) channel
can be parametrized
by the neutrino magnetic moment ($\mu_l$),
often expressed
in units of the Bohr magneton($\mub$).
Its dependence on neutrino energy $\enu$
is given by~\cite{vogelengel}:
\begin{equation}
\label{eq::mm}
\rm{
( \frac{ d \sigma }{ dT } ) _{MS}  ~ = ~
\frac{ \pi \alpha _{em} ^2 {\it \mu_l} ^2 }{ m_e^2 }
 [ \frac{ 1 - T/E_{\nu} }{T} ] ~.
}
\end{equation}
The process can be due to
{\it diagonal} and {\it transition}
magnetic moments, for the cases where
$l_1=l_2$ and $l_1 \neq l_2$, respectively.
The MS interactions involve a flip of spin
and therefore
do not have interference
with the Standard Model (SM) cross-section.
The $\mu_l$ term has a 1/T dependence and hence dominates
at low electron recoil energy over the SM process.

The quantity $\mu_l$ is an effective parameter which,
in the case of $\nuebar$ and large mixings between the
mass eigenstates, can be expressed as~\cite{beacom}:
\begin{equation}
\rm{
\mu_e^2  =  \sum_k \big| \sum_j U_{ej}  \mu_{jk} \big| ^2
}
\end{equation}
where U is the mixing matrix and $\rm{\mu_{jk}}$
are the fundamental constants that characterize
the couplings between the mass eigenstates
$\rm{\nu_j}$ and $\rm{\nu_k}$ with the
electromagnetic field.
The $\nu$-$\gamma$ couplings
probed by $\nu$-e scatterings
is the same as that giving rise to the
neutrino radiative decays~\cite{rdk}:
\begin{equation}
\rm{
\nu_j  \rightarrow   \nu_k   +   \gamma
}
\end{equation}
between $\rm{\nu_j}$ and $\rm{\nu_k}$.
The decay rate $\rm{\Gamma_{jk}}$ is
related to $\rm{\mu_{jk}}$ by:
\begin{equation}
\label{eq::rdk}
\rm{
\Gamma_{jk} ~ = ~ \frac{ ~~ \mu_{jk}^2 ~~ }{ 4 \pi } ~
\frac{ ~~ ( \Delta m_{jk}^2 ) ^3 ~~ }{ m_j^3 } ~
}
\end{equation}
where $\rm{m_j}$ is the mass of $\rm{\nu_j}$
and $\rm{\Delta m_{jk}^2 = m_j^2 - m_k^2 }$.
The total decay rate is given by
\begin{equation}
\rm{
\Gamma_{\nu_e}  =  \sum_k \big| \sum_j U_{ej}  \Gamma_{jk} \big| ^2
}
\end{equation}

Reactor neutrinos provide a sensitive probe
for ``laboratory'' searches  of $\rm{\mu_e}$,
taking advantages of the
high $\nuebar$ flux, low $\enu$ and the better experimental
control via the reactor ON/OFF comparison.
Neutrino-electron scatterings were first
observed in pioneering experiment~\cite{reines}
at Savannah River.
However, a reanalysis of the data by Ref~\cite{vogelengel}
with improved input parameters
on the reactor neutrino spectra 
and $\rm{sin ^2 \theta_W}$
gave a positive signature
consistent with the interpretation of a
finite magnetic
moment at $(2-4) \times 10^{-10}~\mub$.
Other results came
from the Kurtchatov~\cite{kurt} and Rovno~\cite{rovno} experiments
which quoted limits of
$\munuebar$ less than $2.4 \times 10^{-10}~\mub$
and $1.9 \times 10^{-10}~\mub$
at 90\% confidence level (CL), respectively.
However, the lack of experimental details and discussions
in the published work
make it difficult to assess the
robustness of the results.

Theoretically, a minimally-extended
Standard Model would give rise to
$\munu$'s for Dirac neutrinos too small
to be of interest~\cite{smmunu}.
However, there are models~\cite{pdg} 
which can produce large $\munu$ by incorporating
new features like right-handed currents and
transition moments.
Neutrino flavor conversion induced by
resonant or non-resonant spin-flip transitions
in the Sun via its transition magnetic moments
has been considered
as solution to the Solar Neutrino Problem~\cite{munusnuold}.
Stimulated by
the new results from the Super-Kamiokande and SNO experiments,
recent detailed work~\cite{sfpsnunew}
suggested that this scenario is also in excellent
agreement with the existing solar neutrino data.
Alternatively, the measured solar neutrino $\nu_{\odot}$-e spectra has been
used to set limits  of
$\rm{ \mu ^{\odot} _{\nu}  < 1.5 \times 10^{-10} ~ \mub }$
at 90\% CL for the  ``effective'' $\nu_\odot$
magnetic moment~\cite{beacom},
which is in general different
from that of a pure $\nu_l$ state derived
in laboratory experiments.
In addition,
there are astrophysical arguments~\cite{pdg}
from nucleosynthesis,
stellar cooling and SN1987a
which placed limits of the range
$\rm{\mu_e < 10^{-12} - 10^{-13} ~ \mub}$.
However, care should be taken in their
interpretations due to the
model dependence and assumptions
on the neutrino properties
implicit in the derivations.

These discussions show that further laboratory experiments
to put the current limits on more solid grounds and to
improve on the sensitivities are of interest.
Beside the Kuo-Sheng (KS) experiment~\cite{texono}
reported in this article,
there is another on-going experiment MUNU~\cite{munu}
at the Bugey reactor using
a time projection chamber with CF$_4$ gas.

\section{Kuo-Sheng Reactor Neutrino Laboratory}

\begin{figure}
\center
\epsfig{figure=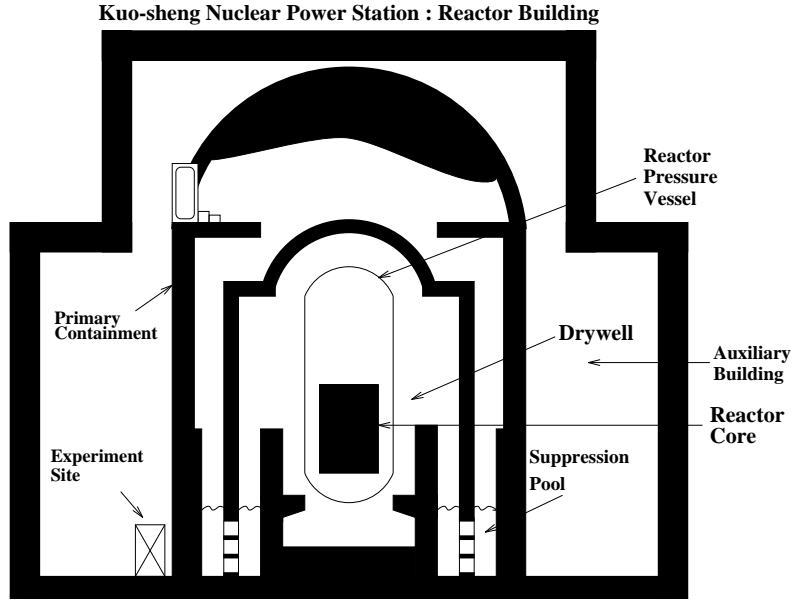,height=8cm,angle=270}
\caption{
Schematic side view, not drawn to scale,
of the Kuo-sheng Nuclear Power Station
Reactor Building,
indicating the experimental site.
The reactor core-detector distance is about
28~m.
}
\label{ksnpssite}
\end{figure}

The ``Kuo-Sheng Reactor Neutrino Laboratory''
is located at a distance of 28~m from the core \#1
of the Kuo-Sheng Nuclear Power Station
at the northern shore of Taiwan~\cite{ksexpt}.
A schematic view is depicted in Figure~\ref{ksnpssite}.

\begin{figure}
\center
\epsfig{figure=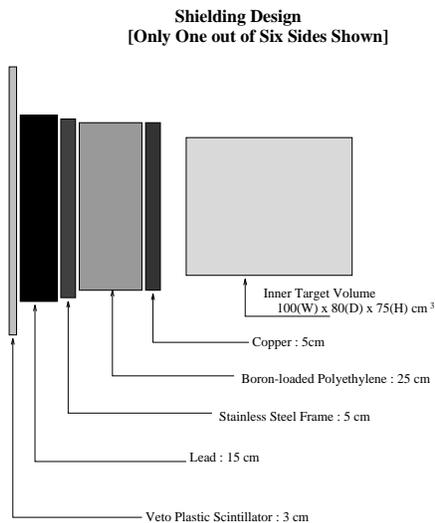,height=2.7in,angle=270}
\caption{
Schematic layout of the inner target space,
passive shieldings and cosmic-ray veto panels.
The coverage is 4$\pi$ but only one face
is shown.
}
\label{shield}
\end{figure}

A multi-purpose ``inner target'' detector space of
100~cm$\times$80~cm$\times$75~cm is
enclosed by 4$\pi$ passive shielding materials
and cosmic-ray veto scintillator panels,
the schematic layout of  which is shown
in Figure~\ref{shield}.
The shieldings provide attenuation
to the ambient neutron and gamma background,
and are made up of, inside out,
5~cm of OFHC copper, 25~cm of boron-loaded
polyethylene, 5~cm of steel and 15~cm of lead.

Different detectors can be placed in the
inner space for the different scientific goals.
The detectors are read out by a versatile
electronics and data acquisition systems~\cite{eledaq}
running on a VME-PCI bus.
Signals are recorded  by 
Flash Analog-to-Digital-Convertor~(FADC) modules
with 20~MHz clock and 8-bit resolution.
The readout allows full recording of all the relevant pulse
shape and timing information for as long as several ms
after the initial trigger.
The reactor laboratory is connected via telephone line to
the home-base laboratory at AS, where remote access 
and monitoring are performed regularly. Data are stored
and accessed in a multi-disks array with a total
of 600~Gbyte memory via IDE-bus in PCs.

It is recognized recently~\cite{jphysg} that
the low energy part of the reactor neutrino
spectra is not well modeled or experimentally
checked. 
Consequently, the uncertainties
induced in the SM $\nuebar$-e cross-sections
can limit the sensitivities of magnetic moment searches
at the domain where
they are comparable or larger than 
the MS interactions.
Therefore, experiments intended for measure 
Standard Model cross sections 
with reactor neutrinos
should focus on higher energies (T$>$1.5~MeV)
while $\munu$ searches should base
on measurements  with T$<$100~keV.

Accordingly, data were taken for Period I Reactor ON/OFF
from June 2001 till April 2002 with these strategies. 
Two detector systems are running in parallel 
using the same data acquisition system 
but independent triggers:
(a) an Ultra Low Background High Purity Germanium (ULB-HPGe),
with a fiducial mass of 1.06 kg,
and (b) 46~kg of CsI(Tl) crystal scintillators.

Preliminary results of the HPGe system is presented in
the following sections. The performance of the
CsI(Tl) modules is discussed elsewhere~\cite{texono,proto}.

\section{Low Background Germanium Detector}

\begin{figure}
\center
\epsfig{figure=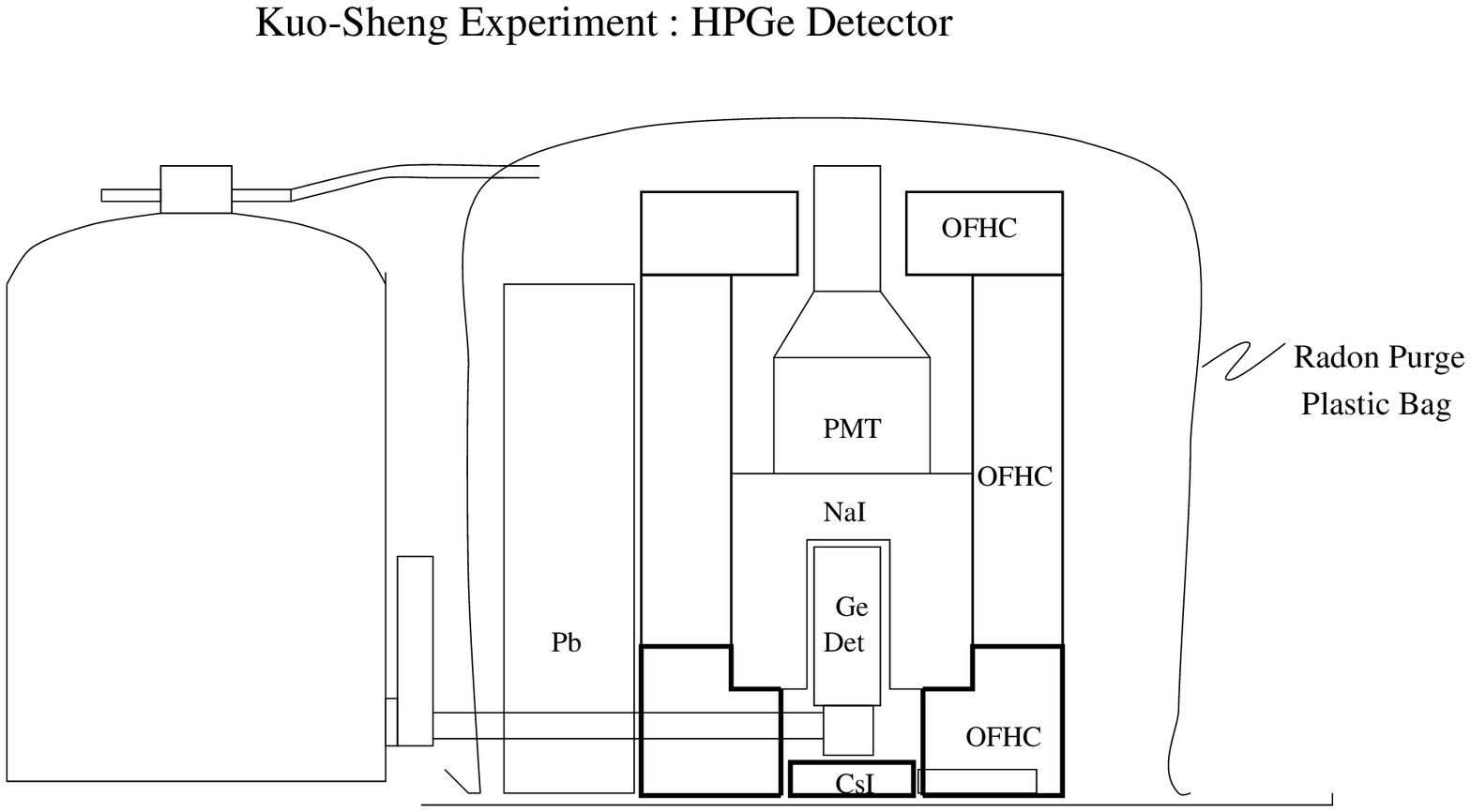,height=12cm,angle=270}
\caption{
Schematic drawings of the ULB-HPGe
detector with its anti-Compton scintillators,
passive shieldings and radon purge system.
}
\label{hpge}
\end{figure}

The low threshold and excellent energy
resolution of the germanium detector make it optimal for
$\munu$ searches~\cite{texono,munuruss}.
The set-up of the KS-Ge experiment
is schematically shown in
Figure~\ref{hpge}.
It is a coaxial germanium detector with an active
target mass of 1.06~kg. The lithium-diffused outer
electrode is 0.7~mm thick. The end-cap cryostat, also
0.7~mm thick, is made of OFHC copper. Both of these
features provide total suppression to ambient $\gamma$-background
below 60~keV, such that events below this energy are either
due to internal activity or external high energy
$\gamma$'s via Compton scattering.
The HPGe was surrounded by an anti-Compton (AC) detector system
made up of two components: (1) an NaI(Tl) well-detector
of thickness 5~cm that fit onto the end-cap cryostat,
the inner wall of which is also made of OFHC copper,
and (2) a 4~cm thick CsI(Tl) detector at the bottom.
Both AC detectors were read out by photo-multipliers (PMTs)
with low-activity glass.
The assembly was surrounded by 3.7~cm of OFHC copper inner
shielding. Another 10~cm of lead provided additions
shieldings on the side of the liquid nitrogen dewar and
pre-amplifier electronics.
The inner shieldings and detectors were covered by a plastic
bag connected to the exhaust line of the dewar, serving
as a purge for the radioactive radon gas.

The HPGe pre-amplifier and AC PMT signals were distributed 
via 10~m cables to
two spectroscopy amplifiers 
at the same 4~$\mu$s shaping time but with different
gain factors.
A very loose amplitude threshold on the SA output
provided the on-line trigger, ensuring all the events
down to the electronics noise edge were recorded.
The SA output, as well as the PMT signals
from the AC detectors,  were recorded by the FADC
for a duration of 25~$\mu$s after the trigger.
The discriminator output of the Veto PMTs were
also recorded.
A {\it random} trigger was provided by an external
clock once per 10~s for sampling the background
level and evaluating the various efficiency factors.
The DAQ system remains active for 2~ms after
a trigger to  record possible time-correlated signatures.
The activities in any part of the detector systems
within 10~$\mu$s prior to the trigger were also recorded.
The typical data taking rate for the HPGe sub-system
was about 1~Hz. The accurately recorded DAQ
dead time was about 10-20~ms per event.

\section{Results}

\begin{figure}
\center
\epsfig{figure=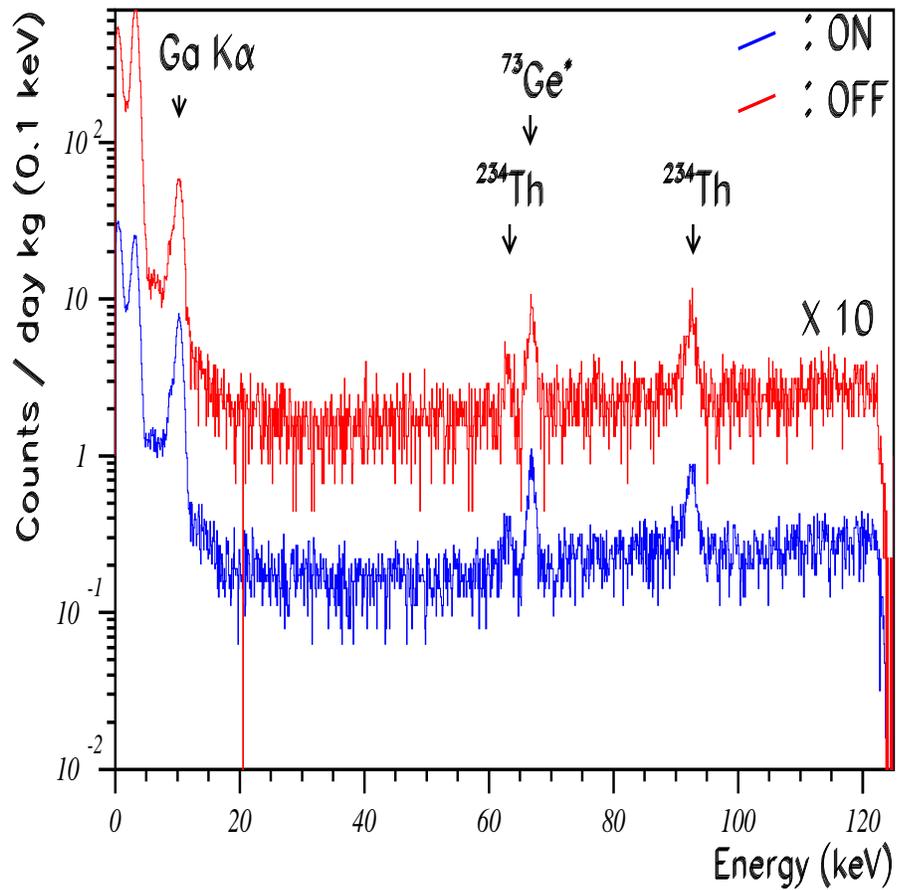,width=12cm}
\caption{
The energy spectra after the anti-Compton
and cosmic veto cuts 
for 60/46 live time days of
Reactor ON/OFF data taking.
The OFF data set is multiplied by 10 for
display purposes.
}
\label{gespec}
\end{figure}

The measured spectra, after cuts of cosmic and
anti-Compton vetos, 
during 60/46~days of reactor ON/OFF data taking are displayed 
in Figure~\ref{gespec}.
The background level of 1~keV$^{-1}$kg$^{-1}$day$^{-1}$
and a detector threshold of 5~keV, comparable to 
those of underground Dark Matter
experiment, are achieved. 
Additional cuts based on pulse shape 
and timing information
are expected to further reduce the background level
and the threshold.

Several lines can be identified:
Ga X-rays at 10.37~keV and $^{73}$Ge$^*$ at 66.7~keV
from internal cosmic-induced activities, and
$^{234}$Th at 63.3~keV and 92.6~keV due to residual ambient
radioactivity from the $\u238$ series in the vicinity of
the target.
The ON and OFF spectra differ in two significant features.
The excess of OFF over ON below the noise
edge of 5~keV is due to instabilities in the trigger threshold
and do not affect the analysis at higher energies.
The excess at the Ga X-ray peaks originates
from the long-lived isotopes
($^{68}$Ge and $^{71}$Ge with half-lives of
271 and 11.4 days, respectively)
activated by cosmic-rays
prior to installation. It
has been checked that the time evolution
of the peak intensity can be fit to
two exponentials consistent with the two known
half-lives.

\begin{figure}
\center
\epsfig{figure=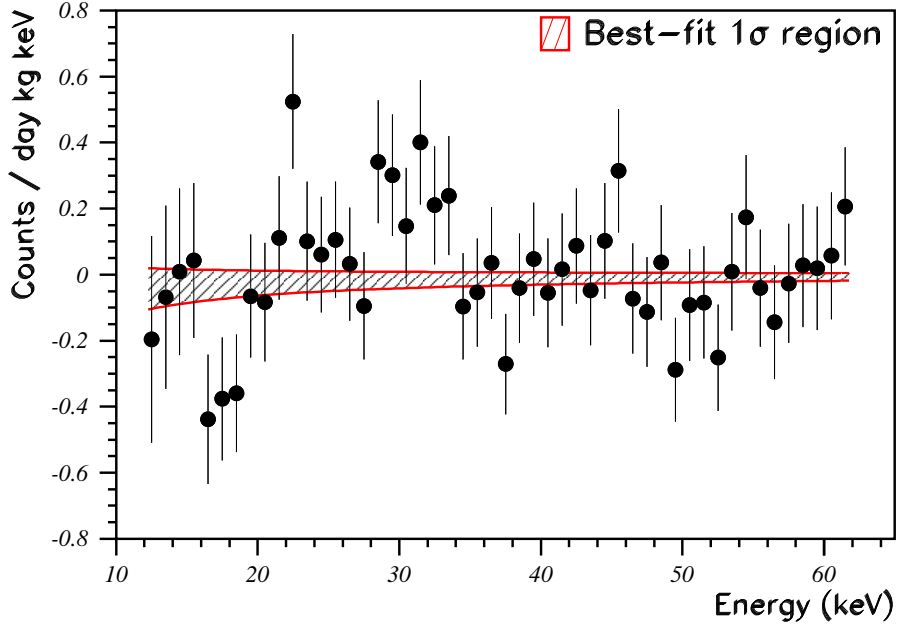,width=12cm}
\caption{
The residual spectrum of the
Reactor ON recoil data
subtracting $\rm{\phi_{OFF}}$.
The best-fit 1$\sigma$ region is also shown.
}
\label{residual}
\end{figure}

The reactor neutrino spectra was evaluated from
reactor operation data using the standard prescriptions
on from fission $\nuebar$~\cite{vogelengel}
together with a low energy contribution due to
neutron capture on $\u238$~\cite{rnule}.
The total flux is $\rm{5.6 \times 10^{12} ~ cm^{-2} s^{-1}}$.
The electron recoil spectra from SM and MS
interactions can then be calculated.
Background in the energy range of 12 to 60~keV are
due to Compton scatterings of higher energy $\gamma$'s.
Accordingly, the Reactor OFF data are fitted 
to a smooth function
\begin{equation}
\rm{
\phi_{OFF} = \alpha_1 e ^{\alpha_2 E} + \alpha_3 + \alpha_4 E  .
}
\end{equation}
A $\chi ^2$/dof of  40/46  is obtained indicating this
background description is valid.
A one-parameter fit is then performed to
the Reactor ON data for 
\begin{equation}
\rm{
\phi_{OFF} +  \phi_{SM} + k^2 \phi_{MS} ~ ,
}
\end{equation}
where
$\rm{ \phi_{SM} }$ and $\rm{\phi_{MS} }$
are the expected recoil spectra
due to weak interactions 
and MS at 
$\munue = 10^{-10} ~ \mub$, respectively.
A best-fit value of
\begin{equation}
\rm{ k^2 = -1.1 \pm 2.5 } 
\end{equation}
at a $\chi ^2$/dof of 55/49 is obtained.
The residual plot and the best-fit 1$\sigma$ region
is depicted in Figure~\ref{residual}.

\begin{figure}
\center
\epsfig{figure=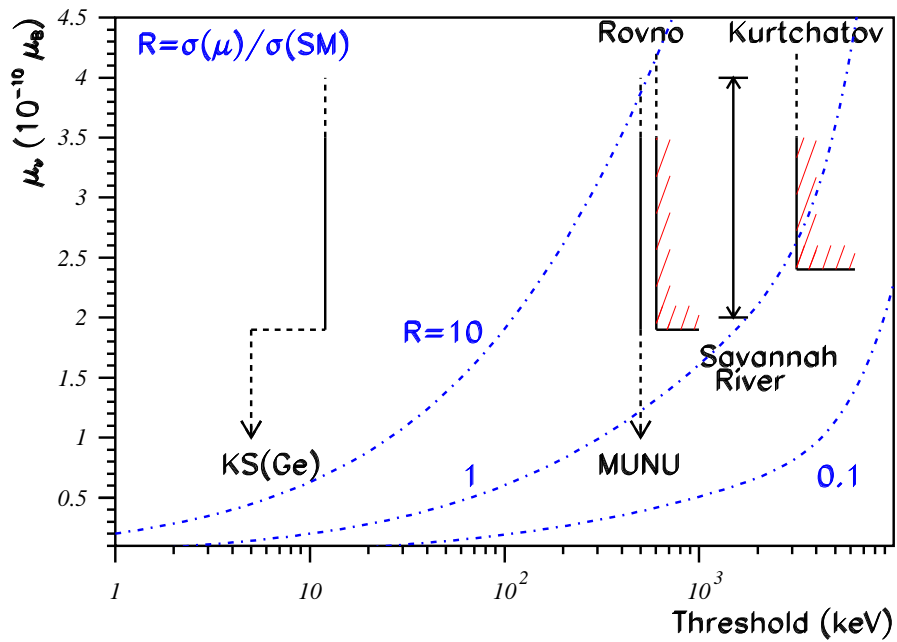,width=12cm}
\caption{
Summary of the results
in the searches of neutrino
magnetic moments with reactor
neutrinos. The dotted line
is the ratio between the cross-sections
due to magnetic moments and Standard Model
weak interactions.
}
\label{munuall}
\end{figure}

A total of 180/60 days of Reactor ON/OFF are taken
for this reactor period.
A limit will be derived when the data analysis is
improved, complete and finalized.
It is expected that the
data would give
world-level sensitivities in the searches of 
$\nuebar$ magnetic moments.
Depicted in Figure~\ref{munuall} is the
summary of all the results in $\munue$ searches
with reactor $\nuebar$ versus the achieved
threshold.
The dotted lines are the
$\rm{\sigma_{\mu}/\sigma_{SM}}$ ratio at a
particular (T,~$\munue$).
The KS experiment operated
at a much lower threshold of 12~keV compared
to previous and current measurements.
The large $\rm{\sigma_{\mu}/\sigma_{SM}}$
ratio at low energy implies
that effects due to
the uncertainties in the SM cross-sections
can be neglected such that the limits
derived are more robust.

\begin{figure}
\center
\epsfig{figure=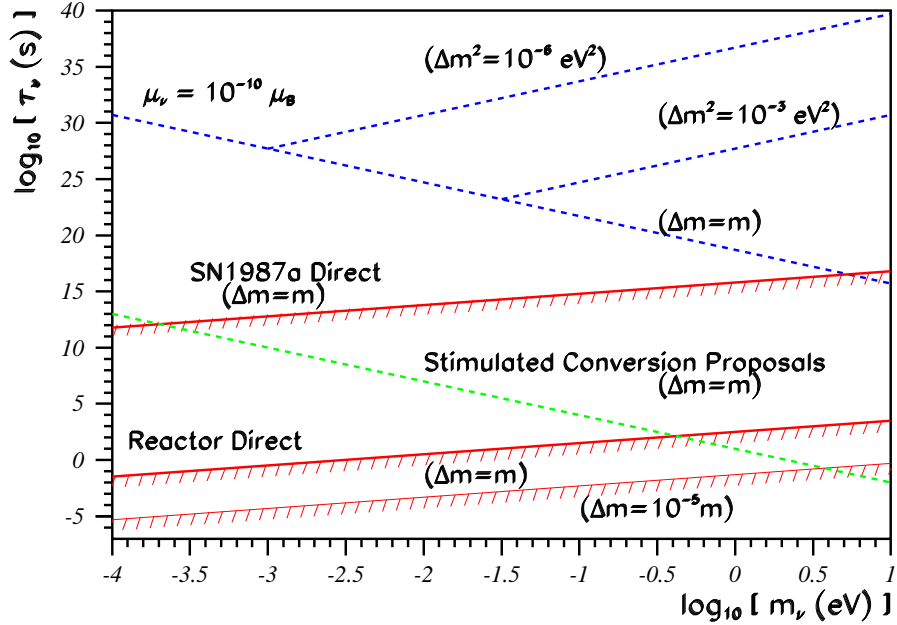,width=12cm}
\caption{
Summary of the results
in the bounds in
radiative decay lifetime of the
neutrino. See text for explanations
}
\label{rdk}
\end{figure}

Indirect bounds on the neutrino radiative decay
rate are inferred using Eq.~\ref{eq::rdk}
and displayed in Figure~\ref{rdk}.
The sensitivity region of $\munu = 10^{-10} \mub$
is shown for illustration purpose.
Superimposed are
the limits from the previous direct searches of excess
$\gamma$'s from reactor neutrinos~\cite{rdkdirect}
and from the supernova SN1987a~\cite{rdksn}.
Also shown is the sensitivity level of proposed
simulated conversion experiments at accelerator~\cite{simcon}.
It can be seen that
$\nu$-e scatterings give much more
stringent bounds than the direct searches.

\section{Summary \& Outlook}

With the strong evidence of physics
beyond the Standard Model revealed in the
neutrino sector,
neutrino physics and astrophysics remains
a central subject in experimental particle physics
in the coming decade and beyond. There
are room for ground-breaking technical innovations -
as well as potentials for surprises in the scientific
results.

A Taiwan, China and U.S.A.
collaboration has been built up 
with the goal of establishing
a experimental program
in neutrino and astro-particle physics.
It is 
the first generation collaborative efforts
in large-scale basic research between scientists
from Taiwan and China.
The flagship effort is to
perform the first-ever particle
physics experiment in Taiwan
at the Kuo-Sheng Reactor Plant.
From the Period I data taking, we expect to achieve
world-level sensitivities in neutrino magnetic
moments and radiative lifetime studies.
A wide spectrum of R\&D projects are 
being pursued. New ideas for future directions
are being explored.

\section{Acknowledgments}

The author is grateful to the scientific members,
technical staff and industrial partners
of TEXONO Collaboration, as well as 
the concerned colleagues
for the invaluable contributions which ``make 
{\it it} happen'' in such a short period
of time.
Funding are
provided by the National Science Council, Taiwan
and 
the National Science Foundation, China, as well
as from the operational funds of the 
collaborating institutes.


\end{document}